\documentclass[twocolumn,showpacs,preprintnumbers,amsmath,amssymb]{revtex4}
\newcommand{\lsim} 
 {\ \raise.35ex\hbox{$<$}\kern-0.75em\lower.5ex\hbox{$\sim$}\ }
\newcommand{\gsim}
 {\ \raise.35ex\hbox{$>$}\kern-0.75em\lower.5ex\hbox{$\sim$}\ }
\def\journal #1#2#3#4{#1 {\bf #2} (#4) #3}

\def\PRB{Phys.\ Rev.\ B}
\def\PRL{Phys.\ Rev.\ Lett.}

\def\IJMP{Int.\ J.\ Mod.\ Phys.}

\def\JMMM{J.~Mag.~Mag.~Mat.}

\def\JPSJ{J.\ Phys.\ Soc.\ Jpn.}

\def\PTP{Prog.\ Theor.\ Phys.}

\usepackage{graphicx}
\usepackage{dcolumn}
\usepackage{bm}
\usepackage{amsmath}
\usepackage{times}

\begin{document}

\title{Predominant Magnetic States in Hubbard Model on Anisotropic 
Triangular Lattices}
\author{ T.~Watanabe$^{1,2}$, H.~Yokoyama$^{3}$, Y.~Tanaka$^{1,2}$, 
and J.~Inoue$^{1}$ }
\affiliation{$^1$Department of Applied Physics, Nagoya University, Nagoya 464-8603, Japan \\
$^2$ CREST Japan Science and Technology Cooperation (JST), Japan \\
$^3$ Department of Physics, Tohoku University, Sendai 980-8578, Japan \\
}
\date{\today}
\begin{abstract}
Using an optimization variational Monte Carlo method, we study the 
half-filled-band Hubbard model on anisotropic triangular lattices, 
as a continuation of the preceding study 
[\journal{\JPSJ}{75}{074707}{2006}]. 
We introduce two new trial states: 
(i) A coexisting state of ($\pi,\pi$)-antiferromagnetic (AF) 
and a $d$-wave singlet gaps, in which we allow for a band 
renormalization effect, and 
(ii) a state with an AF order of 120$^\circ$ spin structure. 
In both states, a first-order metal-to-insulator transition 
occurs at smaller $U/t$ than that of the pure $d$-wave state. 
In insulating regimes, magnetic orders always exist; an ordinary 
($\pi,\pi$)-AF order survives up to $t'/t\sim 0.9$ ($U/t=12$), 
and a 120$^\circ$-AF order becomes dominant for $t'/t\gsim 0.9$. 
The regimes of the robust superconductor and of the nonmagnetic 
insulator the preceding study proposed give way to these magnetic domains. 
\end{abstract}

\pacs{74.70.-b, 74.20.-z}
\maketitle



%
%

%



\section{Introduction\label{sec:intro}}

A series of $\kappa$-(BEDT-TTF)$_2$X [$\kappa$-ET salts] have 
intriguing properties specific to strongly-correlated systems; 
they often undergo unconventional superconductor (SC)-to-insulator 
transitions through the chemical substitution of X or under applied 
pressure, and have good two-dimensionality in conductivity with 
frustrated lattice structure. 
As a model of these compounds, 
the half-filled-band Hubbard model on anisotropic triangular lattices 
\cite{Fukuyama} (an extended square lattice with hopping 
integral $t$ in $x$ and $y$ directions, and $t'$ in one 
diagonal direction [1,1]) has been intensively studied: 
\cite{review} 
%
\begin{eqnarray}
{\cal H}= \sum_{{\bf k}\sigma} \varepsilon_{\bf k} 
   c_{{\bf k}\sigma}^\dag c_{{\bf k}\sigma}+U\sum 
   \limits_i n_{i\uparrow}n_{i\downarrow}, 
\label{eq:model} 
\end{eqnarray}
where 
$
\varepsilon_{\bf k}=-2t(\cos k_x+\cos k_y)-2t'\cos (k_x+k_y),
$
and $U,t,t'>0$. 
To clarify the properties of this model in the strongly-correlated 
region, $U/t\gg 1$, especially Mott transitions, reliable theoretical 
approaches are needed. 
To this end, the present authors recently applied to eq.~(\ref{eq:model}) 
an optimization (or correlated) variational Monte Carlo (VMC) method, 
which can deal with SC and a Mott transition as a continuous function 
of $U/t$. 
Henceforce, we call this preceding study `(I)'.\cite{Wata}
In (I), we chiefly considered various properties of the $d_{x^2-y^2}$-wave 
singlet state, $\Psi_Q^d$, and constructed a ground-state phase 
diagram in the $t'$-$U$ plane, by comparing its energy with that of 
the ordinary ($\pi,\pi$)-antiferromagnetic (AF) state, $\Psi_Q^{\rm AF}$. 
Most of the results are consistent with the behavior of $\kappa$-ET 
salts, but the area of an ($\pi,\pi$)-AF insulator is unexpectedly limited 
($t'/t \lsim 0.4$), in considering the appearance of the AF order 
in e.g. $\kappa$-(ET)$_2{\rm Cu{N(CN)_2}Cl}$ ($t'/t\sim 0.74$), 
as well as the vanishing point of the AF order expected in the $J$-$J'$ 
Heisenberg model ($t'/t\sim 0.8$) \cite{J-J'}. 
As we pointed out in (I), this disagreement possibly stems from the 
fact that the $d$-wave singlet state and the AF state were treated 
separately; thereby, the former state does not include a seed of an AF 
long-range order, and the latter a band renormalization effect. 
\par

For $t'\sim t$, many theoretical studies 
\cite{PIRG,DMFAF,V-CPT,CDMFT,Koretsune} for the Hubbard model have 
obtained results of dominant nonmagnetic insulating state, which are 
consistent with the insulating state found in 
$\kappa$-(ET)$_2$Cu$_2$(CN)$_3$ with $t'/t\sim 1.06$ \cite{Shimizu}. 
Nonetheless, we should be also concerned about the AF order with 
120-degree spin structure, which is considered to prevail in the 
isotropic case of the $J$-$J'$ Heisenberg model.\cite{J-J',tri} 
Actually, a recent VMC study\cite{Weber} for a $t$-$J$-type model
on the isotropic triangular lattice concluded that the 120$^\circ$-AF 
ordered state is dominant in an unexpectedly wide range of doping rate. 
Thus, it is possible that the 120$^\circ$-AF order is robust also 
in the Hubbard model with $t'\sim t$ and sufficiently small values 
of $U/t$ for the organics. 
\par

In this paper, as a continuation of (I), we introduce two trial states: 
(i) A state which includes ($\pi,\pi$)-AF and $d$-wave gaps 
simultaneously;\cite{Giamarchi,Himeda1} and then a band (or Fermi-surface) 
renormalization effect owing to the electron correlation is taken 
into account. \cite{Himeda} 
(ii) A state which exhibits the 120$^\circ$-AF order. 
In addition to these functions, we newly consider SC states with 
pairing symmetries suitable for $t'>t$. 
Our main interest here is the competition among these states 
and those treated in (I). 
It is found that first-order metal-to-insulator transitions always 
occur at smaller values of $U/t$ than those for the pure $d$-wave state. 
In the insulating regime, the ($\pi,\pi$)-AF order remains up to 
$t'/t\sim 0.9$, 
owing to the band renormalization effect we considered in the 
coexisting state, and the 120$^\circ$-AF order becomes predominant 
in the range of $t'/t\gsim 0.9$. 
Consequently, a magnetic order, namely the ($\pi,\pi$)-AF or 120$^\circ$-AF 
order, always exists in the insulating regime, and a regime of 
a nonmagnetic insulator vanishes. 
In addition, a domain of dominant SC found in (I) disappears 
within the present results. 
The previous phase diagram is substantially modified. 
\par

In \ref{sec:method}, we explain the trial wave functions used, 
and recapitulate the main points of (I) as a motivation of this study. 
In \ref{sec:results}, we represent the VMC results. 
In \ref{sec:conclusion}, we briefly summarize this study, 
and compare with experimental and other theoretical results. 
\par

A part of the results have been reported before.\cite{ICM}

\section{Wave functions\label{sec:method}}

As usual, we use Jastrow-type trial wave functions: $\Psi={\cal P}\Phi$, 
in which $\Phi$ denotes a one-body (Hartree-Fock) part expressed 
as a Slater determinant, and ${\cal P}$ a many-body correlation 
factor. 
In \ref{sec:cor}, we describe the correlation factor ${\cal P}$. 
In \ref{sec:coexist}, we point out insufficient points in the 
wave functions used in (I), and introduce a coexisting state 
of the ($\pi,\pi$)-AF and $d$-wave gaps in which the one-body band 
structure is modified by optimizing a hopping parameter $\tilde t'$, 
as renormalization owing to electron correlation. 
In \ref{sec:120}, we formulate a state with an AF order of 120$^\circ$ 
spin structure, as an another new trial state. 
In \ref{sec:VMC}, we briefly touch on the conditions of the VMC 
calculations. 

\subsection{Correlation factor \label{sec:cor}}

When one treats the Hubbard model on the basis of a variational method, 
it is crucial to introduce, in addition to the well-known Gutzwiller 
(onsite) factor ${\cal P}_{\rm G}$, \cite{Gutz,YS1} intersite correlation 
factors \cite{Kaplan,YS3} into Jastrow-type wave functions. 
In particular, near half filling, the binding effect of a doubly-occupied 
site (doublon) to an empty site (holon) is indispensable to describe 
a Mott transition as well as various quantities appropriately. 
\cite{YS3}
To this end, we have repeatedly studied \cite{YokoPTP,YTOT,YOT,Wata} 
a four-body factor formally written as, 
\begin{eqnarray}
{\cal P}_Q=
\prod_i{\bigl(1-\mu Q_i^{\tau} \bigr)
\bigl(1-\mu'Q_i^{\tau'}\bigr)}, 
\label{eq:PQ}
\end{eqnarray}
\begin{eqnarray}
Q_i^{\tau(\tau')}=\prod_{\tau(\tau')} 
\left[d_i(1-e_{i+\tau(\tau')})+e_i(1-d_{i+\tau(\tau')})\right], 
\label{eq:Qi}
\end{eqnarray} 
in which $d_i=n_{i\uparrow}n_{i\downarrow}$, 
$e_i=(1-n_{i\uparrow})(1-n_{i\downarrow})$, 
and $\tau$ ($\tau'$) runs over all the adjacent sites in the bond 
directions of $t$ ($t'$). 
In eq.~(\ref{eq:PQ}), $\mu$ ($\mu'$) is a variational parameter which 
controls the binding strength between a doublon and a holon in the 
bond direction $t$ ($t'$). 
We have confirmed that ${\cal P}_Q$ works effectively in the model,
eq.~(\ref{eq:model}).\cite{Wata}
\par

\subsection{Coexisting state of $d$-wave and AF gaps \label{sec:coexist}}

Using ${\cal P}={\cal P}_Q{\cal P}_{\rm G}$, 
we mainly studied, in (I), a $d_{x^2-y^2}$-wave singlet state: 
$\Psi_Q^d={\cal P}\Phi_d$, where $\Phi_d$ is the BCS function with 
a $d_{x^2-y^2}$-wave gap: 
\begin{equation}
\Delta_{\bf k}=\Delta_d(\cos{k_x}-\cos{k_y}). 
\label{eq:d-gap}
\end{equation}
In $\Psi_Q^d$, we allow for renormalization of the one-body band 
$\varepsilon_{\bf k}$ owing to electron correlation, by varying 
$t' (\equiv\tilde t')$ in $\Phi_d$ as a variational parameter, \cite{Himeda}
independently of $t'$ fixed in the Hamiltonian eq.~(\ref{eq:model}). 
In (I), we obtained the following results within $\Psi_Q^d$. 
{\bf (i)} A first-order Mott (conductor-to-nonmagnetic-insulator) transition 
takes place for arbitrary $t'/t$ at $U=U_{\rm c}$ roughly of the bandwidth. 
This transition is induced by the binding (and unbinding) of a doublon 
(negatively charged) to a holon (positively charged), unlike the famous 
Brinkman-Rice transition. \cite{BR} 
{\bf (ii)} Robust $d$-wave SC appears in a restricted parameter range 
immediately below $U_{\rm c}$ and of weak frustration ($t'/t \lsim 0.7$). 
This SC is considered to be induced by a short-range ($\pi,\pi$)-AF spin 
correlation, because whenever the superconducting (SC) correlation function 
is sizably enhanced, the spin structure factor $S({\bf q})$ has a sharp peak 
at the AF wave number, ${\bf q}={\bf K}=(\pi,\pi)$. 
{\bf (iii)} In the insulating regime, $\Psi_Q^d$ exhibits a spin-gap behavior 
and does not have an ($\pi,\pi$)-AF long-range order, 
although $S({\bf q})$ has a sharp peak at ${\bf q}={\bf K}$, 
namely a short-range AF correlation considerably develops. 
\par

\begin{figure}[!t]
\begin{center}
\includegraphics[width=7.5cm,clip]{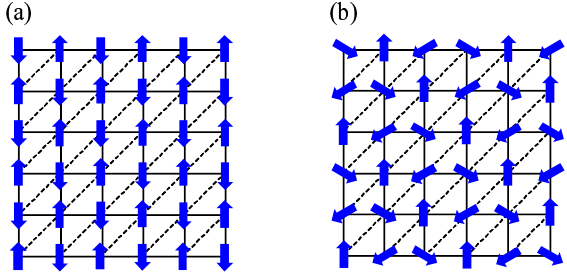}
\end{center}
\caption{(Color online) 
Schematic representation of spin structure in two AF orders studied 
in this paper for the anisotropic triangular lattice: 
(a) an ordinary ($\pi,\pi$)-AF order, and 
(b) an AF order with 120$^\circ$ spin structure. 
}
\label{fig:twoAF}
\end{figure}

To consider the competition between $\Psi_Q^d$ and a state with the 
($\pi,\pi$)-AF long-range order [see Fig.~\ref{fig:twoAF}(a)], 
which should prevail for small $t'/t$, we also studied a projected AF state, 
$\Psi_Q^{\rm AF}={\cal P}\Phi_{\rm AF}$, where $\Phi_{\rm AF}$ 
is a mean-field-type ($\pi,\pi$)-AF state. 
In $\Psi_Q^{\rm AF}$, we did not renormalize $\tilde t'$, because the 
variational energy $E$ to be minimized becomes a discrete 
function of $\tilde t'/t$. 
We found that {\bf (iv)} the stable range of $\Psi_Q^{\rm AF}$ against 
$\Psi_Q^d$ is restricted to a weakly frustrated regime, 
$t'/t\lsim 0.4$ (for $U/t=6$), and this range tends to shrink as 
$U/t$ increases. 
As notified in (I), the above results (iii) and (iv) are not consistent 
with various approximate results \cite{J-J'} for the corresponding 
$J$-$J'$ spin model, which predict that the ($\pi,\pi$)-AF domain 
continues up to $t'/t\sim 0.8$. 
To resolve this disagreement, a seed of the AF order should be introduced 
into $\Psi_Q^d$, and the renormalization of $\varepsilon_{\bf k}$ owing 
to $U$ into $\Psi_Q^{\rm AF}$.
\par

In this paper, we study a wave function, 
$\Psi_Q^{\rm co}={\cal P}\Phi_{\rm co}$, which meets the above 
requirements by merging 
$\Psi_Q^d$ and $\Psi_Q^{\rm AF}$. 
In $\Psi_Q^{\rm co}$, the $d$-wave gap and an AF order can coexist. 
\cite{Giamarchi} 
The one-body part is written as, 
\begin{equation}
\Phi_{\rm co} =\left(\sum_{\bf k}\varphi_{\bf k}
b_{{\bf k},\uparrow}^\dagger b_{{\bf -k},\downarrow}^\dagger
\right)^{N_{\rm e}/2}|0\rangle, 
\end{equation}
in which $N_{\rm e}$ is the electron number, and $\varphi_{\bf k}$ is 
the ratio of BCS coefficients: 
\begin{equation}
\varphi_{\bf k}=\frac{v_{\bf k}}{u_{\bf k}}=
\frac{\Delta_{\bf k}}
{\tilde\varepsilon_{\bf k}-\zeta+
\sqrt{(\tilde\varepsilon_{\bf k}-\zeta)^2+
\Delta_{\bf k}^2}}, 
\label{eq:BCSDelta}
\end{equation}
with 
\begin{equation}
\tilde\varepsilon_{\bf k}=-2t(\cos k_x+\cos k_y)-2\tilde t'\cos (k_x+k_y), 
\end{equation}
and $b^\dagger$ is a creation operator that diagonalizes the ordinary 
($\pi,\pi$)-AF Hartree-Fock Hamiltonian, and is given as 
\begin{eqnarray}
b_{{\bf k},\sigma}^\dagger&=&\alpha_{\bf k} c_{{\bf k},\sigma}^\dagger 
+ \varsigma \beta_{\bf k} c_{{\bf k}+{\bf K},\sigma}^\dagger, \\
b_{{\bf k+K},\sigma}^\dagger&=&-\varsigma \beta_{\bf k} 
c_{{\bf k},\sigma}^\dagger + 
\alpha_{\bf k} c_{{\bf k}+{\bf K},\sigma}^\dagger, 
\end{eqnarray}
\begin{equation}
\alpha_{\bf k}\left(\beta_{\bf k}\right) = 
\sqrt{ {\frac{1}{2}\left( {1-(+)\frac{{\gamma_{\bf k}}}
{{\sqrt {\gamma_{\bf k}^2+\Delta_{{\rm AF}}^2 } }}} \right)}}, 
\end{equation}
with $\gamma_{\bf k} = -2t(\cos k_x + \cos k_y)$ and 
$\varsigma=+(-)1$, according as $\sigma=\ \uparrow$ ($\downarrow$).
In addition to the six parameters in $\Psi_Q^d$, namely, $g$ 
[Gutzwiller (onsite) parameter], $\mu$, $\mu'$, $\Delta_d$, $\zeta$ 
(chemical potential) and $\tilde t'$, 
$\Psi_Q^{\rm co}$ has the seventh parameter $\Delta_{\rm AF}$, which 
controls the staggered spin field and is closely connected to the 
($\pi,\pi$)-AF order parameter $m_{\rm s}$ (sublattice magnetization). 
Note that, in contrast to $\Delta_{\rm AF}$, a finite optimized 
value of $\Delta_d$ does not necessarily mean that a SC gap opens, but an 
insulating spin gap. 
For $\Delta_{\rm AF}\ (\Delta_d)\rightarrow 0$, $\Psi_Q^{\rm co}$ 
is reduced to $\Psi_Q^d$ ($\Psi_Q^{\rm AF})$. 
Thus, we may regard $\Psi_Q^{\rm co}$ as $\Psi_Q^d$ in which the 
($\pi,\pi$)-AF long-range order can arise, and also as $\Psi_Q^{\rm AF}$ 
into which a band renormalization effect is introduced through 
the $d$-wave gap. 
\par

\subsection{AF-ordered state with 120-degree spin structure 
\label{sec:120}}

As discussed in \ref{sec:intro}, an AF-ordered state with 120$^\circ$ 
spin structure [see Fig.~\ref{fig:twoAF}(b)] is plausible for the region of $t'/t\sim 1$. 
We introduce such a state, $\Psi_{\rm 120}={\cal P}\Phi_{\rm 120}$, 
for the Hubbard model eq.~(\ref{eq:model}), and check its stability 
for finite values of $U/t$ and consistency with the results obtained 
for $U/t=\infty$.  \cite{J-J',tri}
\par

\begin{figure}[!t] 
\begin{center}
\includegraphics[width=7.7cm,clip]{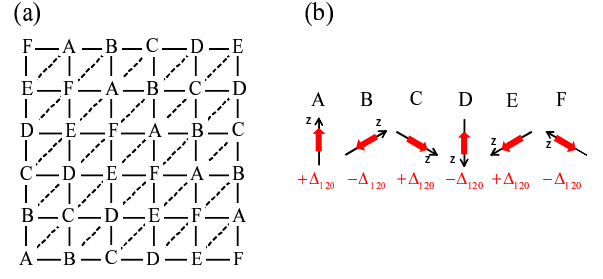}
\end{center}
\caption{(Color online) 
Schematic explanation of Hartree-Fock approximation for an AF order 
with 120$^\circ$ spin structure. 
In (a), it is shown how we divide the anisotropic triangular lattice 
into six sublattices (A-F) with different directions of a spin quantization 
axis, which are illustrated in (b): 
the axis of B (C,D,E,F,A) sublattice is obtained by turning that 
of A (B,C,D,E,F) sublattice by 60 degrees. 
For these sublattices, we suppose that the gap parameter is staggered, 
namely $\Delta_{\rm 120}$, $-\Delta_{\rm 120}$, $\Delta_{\rm 120},\cdots$, 
leading to the formation of a 120$^\circ$-AF order in Fig.~\ref{fig:twoAF}(b). 
}
\label{fig:120de}
\end{figure}

As the one-body part, $\Phi_{\rm 120}$, we use a Hartree-Fock 
ground state for the Hamiltonian eq.~(\ref{eq:model}). 
As explained in Fig.~\ref{fig:120de}, we consider six sublattices 
(A-F); the spin quantization axis of a sublattice is turned by 
60 degrees from that of a neighboring sublattice. 
Using this scheme, the Hamiltonian eq.~(\ref{eq:model}) is 
transformed to 
$$
\arraycolsep=0pt %
H = -\sum\limits_{\lambda}
{\left[ {t
\sum\limits_{<i_\lambda,j_{\lambda+1}>} 
{\left( {\begin{array}{*{20}c} 
{a_{i_\lambda,\uparrow}^\dag} & {a_{i_\lambda,\downarrow}^\dag} \\
\end{array}} \right) 
R\left(\frac{\pi}{6}\right) 
\left( {\begin{array}{*{20}c} 
{a_{j_{\lambda+1},\uparrow} }  \\
{a_{j_{\lambda+1},\downarrow} }  \\
\end{array}} \right)} } \right.} 
$$
$$
\arraycolsep=0pt %
+ \left. {t'\sum\limits_{(i_\lambda,j_{\lambda+2})} 
{\left( {\begin{array}{*{20}c}
{a_{i_\lambda,\uparrow}^\dag} & {a_{i_\lambda,\downarrow}^\dag} \\
\end{array}} \right)
R\left(\frac{\pi}{3}\right) 
\left( {\begin{array}{*{20}c} 
{a_{j_{\lambda+2},\uparrow} }  \\
{a_{j_{\lambda+2},\downarrow} }  \\
\end{array}} \right)} } \right] + {\rm h.c.} 
$$
\begin{eqnarray}
+ U\sum\limits_\lambda {\sum\limits_{i_\lambda} 
{n_{i\uparrow}^T n_{i\downarrow}^T} }, 
\label{eq:TH}
\end{eqnarray}
where 
\begin{equation}
R(\theta ) = \left( {\begin{array}{*{20}c}
   {\cos (\theta )} & { - \sin (\theta )}  \\
   {\sin (\theta )} & {\cos (\theta )}  \\
\end{array}} \right), 
\end{equation}
$a^\dagger_{i,\sigma}$ is a creation operator in the sublattice 
representation, $n_{i\sigma}^T = a_{i,\sigma}^\dag a_{i,\sigma}$, 
$\lambda$ ($=$A-F) is a sublattice index, $i_\lambda$ runs over 
all the sites on sublattice $\lambda$, and an angle (round) bracket 
in the summation indices in eq.~(\ref{eq:TH}) indicates 
a nearest(diagonal)-neighbor pair. 
We apply a Hartree-Fock decoupling to the interaction term in 
eq.~(\ref{eq:TH}), 
\begin{equation}
\sum_i {U n_{i\uparrow}^T n_{i\downarrow}^T} 
\sim 
\sum_i {U\left( {\left\langle {n_{i\uparrow}^T} 
\right\rangle n_{i\downarrow}^T + \left\langle {n_{i\downarrow}^T} 
\right\rangle n_{i\uparrow}^T } \right)}
+ {\rm const.} ,
\end{equation}
and assume that the gap is staggered as 
\begin{equation}
\frac{U}{2}\left(\langle n_{i_\lambda   \uparrow }^T\rangle  
                -\langle n_{i_\lambda   \downarrow }^T\rangle \right)
\equiv \left\{{\begin{array}{*{20}c}
   { + \Delta _{120} {\rm{~~~if~}}\lambda  = {\rm{A,C,E}}}  \\
   { - \Delta _{120} {\rm{~~~if~}}\lambda  = {\rm{B,D,F}}}  \\
\end{array}} \right., 
\end{equation}
to form a 120$^\circ$-AF order. 
Using the operators for sublattices, the Hartree-Fock Hamiltonian 
in the wave-number representation is given as, 
$$
\arraycolsep=1pt %
H_{\rm HF} = \sum\limits_{{\bf{k}},\sigma } 
{\left( {\begin{array}{*{20}c}
        {\begin{array}{*{20}c}
        {\begin{array}{*{20}c}
   {a_{{\bf{k}},\sigma}^{\dag A}} & {a_{{\bf{k}},\sigma}^{\dag B}} \\
   \end{array}} 
 & {a_{{\bf{k}},\sigma}^{\dag C}} & {a_{{\bf{k}},\sigma}^{\dag D}} \\
   \end{array}} 
 & {a_{{\bf{k}},\sigma}^{\dag E}} & {a_{{\bf{k}},\sigma}^{\dag F}} \\
   \end{array}} \right)} 
$$
%
$$
\arraycolsep=-2pt %
\times 
\left( {\begin{array}{*{20}c}
   {-\sigma \Delta _{120}} & {A_1} & {A_2^*} & 0 & {A_2} & {A_1^*}  \\
   {A_1^*} & {\sigma \Delta_{120}} & {A_1} & {A_2^*} & 0 & {A_2}  \\
   {A_2} & {A_1^*} & {-\sigma \Delta_{120}} & {A_1} & {A_2^*} & 0  \\
   0 & {A_2} & {A_1^*} & {\sigma \Delta_{120}} & {A_1} & {A_2^*}  \\
   {A_2^*} & 0 & {A_2} & {A_1^*} & {-\sigma \Delta_{120}} & {A_1}  \\
   {A_1} & {A_2^*} & 0 & {A_2} & {A_1^*} & {\sigma \Delta_{120}}  \\
\end{array}} \right)
\arraycolsep=1pt %
\left( {\begin{array}{*{20}c}
   {\begin{array}{*{20}c}
   {a_{{\bf{k}},\sigma}^A}  \\
   {a_{{\bf{k}},\sigma}^B}  \\
   {a_{{\bf{k}},\sigma}^C}  \\
\end{array}}  \\
   {a_{{\bf{k}},\sigma}^D}  \\
   {a_{{\bf{k}},\sigma}^E}  \\
   {a_{{\bf{k}},\sigma}^F}  \\
\end{array}} \right)
$$
%
$$
\arraycolsep=1pt %
+ \sum\limits_{{\bf{k}}\sigma } 
{\left( {\begin{array}{*{20}c}
        {\begin{array}{*{20}c}
        {\begin{array}{*{20}c}
   {a_{{\bf{k}},\sigma}^{\dag A}} & {a_{{\bf{k}},\sigma}^{\dag B}} \\
   \end{array}} 
 & {a_{{\bf{k}},\sigma}^{\dag C}} & {a_{{\bf{k}},\sigma}^{\dag D}} \\
   \end{array}} 
 & {a_{{\bf{k}},\sigma}^{\dag E}} & {a_{{\bf{k}},\sigma}^{\dag F}} \\
   \end{array}} \right)} 
$$
%
$$
\arraycolsep=4pt %
\times
\left( {\begin{array}{*{20}c}
   0 & {B_{1+}} & {B_{2+}} & 0 & {B_{2-}} & {B_{1-}}  \\
   {B_{1-}} & 0 & {B_{1+}} & {B_{2+}} & 0 & {B_{2-}}  \\
   {B_{2-}} & {B_{1-}} & 0 & {B_{1+}} & {B_{2+}} & 0  \\
   0 & {B_{2-}} & {B_{1-}} & 0 & {B_{1+}} & {B_{2+}}  \\
   {B_{2+}} & 0 & {B_{2-}} & {B_{1-}} & 0 & {B_{1+}}  \\
   {B_{1+}} & {B_{2+}} & 0 & {B_{2-}} & {B_{1-}} & 0  \\
\end{array}} \right)
\arraycolsep=1pt %
\left( {\begin{array}{*{20}c}
   {\begin{array}{*{20}c}
   {a_{{\bf{k}},-\sigma}^A}  \\
   {a_{{\bf{k}},-\sigma}^B}  \\
   {a_{{\bf{k}},-\sigma}^C}  \\
\end{array}}  \\
   {a_{{\bf{k}},-\sigma}^D}  \\
   {a_{{\bf{k}},-\sigma}^E}  \\
   {a_{{\bf{k}},-\sigma}^F}  \\
\end{array}} \right) 
$$
\begin{eqnarray}
+ {\rm const.}, 
\label{eq:kH}
\end{eqnarray}
where $a_{{\bf{k}},\sigma}^{\lambda\dag}$ is the Fourier transformation of 
$a^\dagger_{i\lambda,\sigma}$, and
\begin{eqnarray}
A_1 &=& -t\cos(\pi /6)(e^{-ik_x} + e^{-ik_y}), \nonumber\\
A_2 &=& -t'\cos(\pi/3)e^{-i(k_x+k_y)}, \nonumber\\
B_{1+} &=&  t\sin(\pi/6)(e^{-ik_x} + e^{-ik_y}), \nonumber\\
B_{1-} &=& -t\sin(\pi/6)(e^{ ik_x} + e^{ ik_y}), \nonumber\\
B_{2+} &=& t'\sin(\pi/3)e^{-i(k_x+k_y)}, \nonumber\\
B_{2-} &=& -t'\sin(\pi/3)e^{ i(k_x+k_y)}. 
\end{eqnarray}
As $\Phi_{\rm 120}$, we adopt the lowest-energy eigenvector 
obtained by diagonalizing eq.~(\ref{eq:kH}). 
However, we do not determine $\Delta_{\rm 120}$ by a self-consistent 
equation in the Hartree-Fock approximation, 
but optimize $\Delta_{\rm 120}$ as a variational parameter 
in $\Psi_Q^{\rm 120}$ simultaneously with the other parameters 
with respect to the original Hamiltonian eq.~(\ref{eq:TH}). 
If $\Delta_{\rm 120}$ is finite, all sublattices have 
staggered spin densities, constituting the 120$^\circ$ spin structure. 
\par

\subsection{\label{sec:VMC}Variational Monte Carlo calculations}

Generally, it is not easy to accurately calculate expectation values 
of a many-body wave function with analytic approaches. 
Here, we apply an optimization VMC method,\cite{Umrigar} which 
effectively minimizes the variational energy and makes a
virtually accurate evaluation, to the wave functions mentioned 
in this section. 
We have performed VMC calculations mainly for the lattice of 
$N_{\rm s}=L\times L$ sites with $L=10$ and 12. 
The conditions of calculations here are mostly the same as those 
in (I). 

\section{Results\label{sec:results}}

In \ref{sec:ene}, we consider the energies of $\Psi_Q^{\rm co}$ 
and $\Psi_Q^{\rm 120}$, 
and the critical behaviors appearing in them. 
In \ref{sec:transition}, we show these critical behaviors indicate 
a metal-to-insulator transition. 
In \ref{sec:behavior}, we discuss the properties of the AF order 
in the insulating regime of $\Psi_Q^{\rm co}$, and 
the eventual phase diagram. 
In \ref{sec:extd}, we consider the BCS state with another pairing 
symmetries expected for the region of $t'\gsim t$. 
\par

\subsection{\label{sec:ene}Stability of coexisting state and 
120$^\circ$-AF state}
\begin{figure}[!t]
\begin{center}
\includegraphics[width=8.0cm,clip]{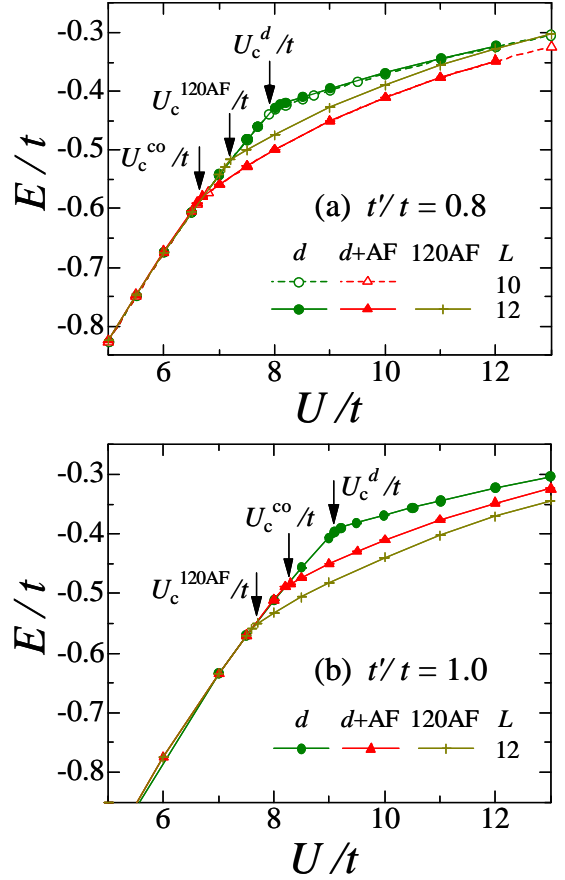}
\end{center}
\vspace{-0.5cm}
\caption{(Color online) 
Total energies of the coexisting state $\Psi_Q^{\rm co}$ [$d$+AF], 
the 120$^\circ$-AF state $\Psi_Q^{\rm 120}$ [120AF], 
and the $d$-wave state $\Psi_Q^d$ [$d$] are compared as a function 
of the correlation strength, for (a) $t'/t=0.8$ and (b) 1.0. 
The critical values of Mott transitions $U_{\rm c}/t$ are indicated 
by arrows for respective states. 
Although the data for $L=10$ and $12$ are plotted, the system-size 
dependence is almost negligible in this scale. 
For $t'/t=1$, the system of $L=10$ is not used because the 
closed-shell condition is not satisfied. 
}
\label{fig:ene}
\end{figure}
%
\begin{figure}[!t]
\begin{center}
\includegraphics[width=8.5cm,clip]{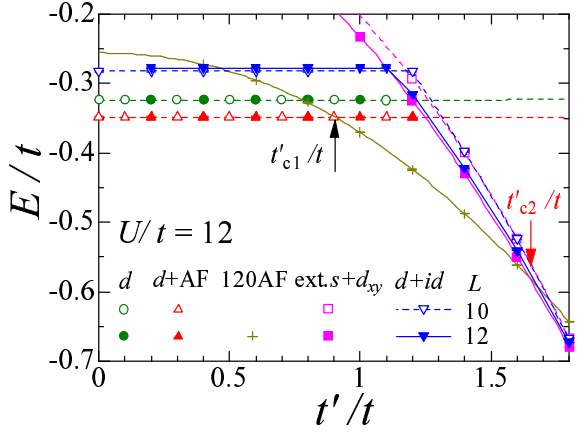}
\end{center}
\vspace{-0.5cm}
\caption{(Color online) 
Comparison of total energies in the insulating regime ($U/t=12$) 
as a function of $t'/t$ among various states: 
a coexisting state $\Psi_Q^{\rm co}$ [$d$+AF], 
a 120$^\circ$-AF state $\Psi_Q^{\rm 120}$ [120AF] and 
three singlet states: 
a $d$ wave $\Psi_Q^d$ [$d$], 
an ${\rm ext.}s$+$d_{xy}$ wave $\Psi_Q^{s+d'}$ [${\rm ext.}s$+$d_{xy}$] and 
a $d_{x^2-y^2}$+$id_{xy}$ wave $\Psi_Q^{d+id}$ [$d$+$id$]. 
The latter two states will be discussed in \ref{sec:extd}. 
The arrows indicate the boundary values between $t'_{\rm c}/t$ and 
$t'_{\rm c2}/t$ satisfying $E^{\rm 120}=E^{\rm co}$ and 
$E^{\rm 120}=E^{s+d'}$, respectively. 
}
\label{fig:enetp}
\end{figure}
%
We start with the energy reduction of the coexisting state 
$\Psi_Q^{\rm co}$ and the 120$^\circ$-AF state $\Psi_Q^{\rm 120}$ 
for $t'\sim t$. 
In Figs.~\ref{fig:ene}(a) and \ref{fig:ene}(b), the total energy 
per site $E$ is compared among $\Psi_Q^{\rm co}$ ($E^{\rm co}$), 
$\Psi_Q^{\rm 120}$ ($E^{\rm 120}$) and $\Psi_Q^d$ ($E^d$) 
for $t'/t=0.8$ and 1.0, respectively. 
For both values of $t'/t$, the curves of $E/t$ for the three states 
are almost indistinguishable from one another for small $U/t$, 
whereas they separate with cusps as $U/t$ becomes large. 
In fact, as we will see shortly, these cusps indicate metal-insulator 
transitions. 
For $t'/t=0.8$, $E^{\rm co}$ exhibits a cusp first 
at $U=U_{\rm c}^{\rm co}=6.65t\pm 0.05t$ and becomes appreciably lower 
than both $E^d$ and $E^{\rm 120}$ for $U>U_{\rm c}^{\rm co}$. 
On the other hand, for $t'/t=1.0$, $E^{\rm 120}$ exhibits a cusp first 
at $U=U_{\rm c}^{\rm 120}=7.65t\pm 0.05t$ and becomes the lowest for 
$U>U_{\rm c}^{\rm 120}$. 
Thus, the lowest-energy state for large $U/t$ is switched from 
$\Psi_Q^{\rm co}$ to $\Psi_Q^{\rm 120}$ in the range of $0.8<t'/t<1.0$. 
To see $t'/t$ dependence of $E/t$ in the insulating regime ($U>U_{\rm c}$), 
we plot the total energies at $U/t=12$ of various states in 
Fig.~\ref{fig:enetp}. 
For $t'<t'_{\rm c}\sim 0.90t$, the coexisting state is the most stable, 
and the decrease in $E/t$ from $E^d/t$ estimated in (I) is 
approximately 7.6\%, irrespective of the value of $t'/t$. 
This invariant behavior of $E/t$ with respect to $t'/t$ is caused 
by marked band renormalization; this point will be discussed 
in detail in \ref{sec:behavior}. 
In contrast, $E^{\rm 120}$ decreases rapidly as $t'/t$ increases, 
and becomes the lowest for $t'>t'_{\rm c1}$. 
As expected, $\Psi_Q^{\rm 120}$ becomes predominant near 
the symmetric point ($t'/t\sim 1$). 
Consequently, the area where the pure $d$-wave singlet state 
$\Psi_Q^d$ prevails does not appear in the insulating regime. 
\par

\begin{figure}[!t]
\begin{center}
\includegraphics[width=7.0cm,clip]{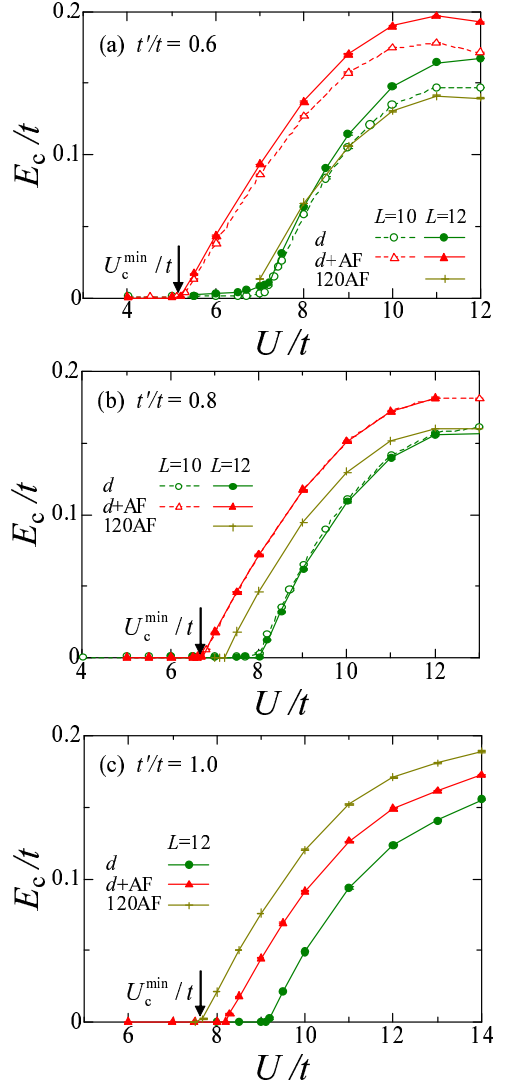}
\end{center}
\vspace{-0.8cm}
\caption{(Color online)
Comparison of the condensation energy $E_{\rm c}/t$ 
among $\Psi_Q^{\rm co}$ ($d$+AF), $\Psi_Q^{\rm 120}$ (120AF) and 
$\Psi_Q^d$ ($d$), for (a) $t'/t=0.6$, (b) 0.8 and (c) 1.0. 
The arrow on the horizontal axis in each panel indicates the critical 
point of the metal-insulator transition arising at the smallest 
$U_{rm c}$ ($\equiv U_{rm c}^{\rm min}$) among those states. 
}
\label{fig:cond}
\end{figure}
%
To discuss the energy reduction more closely, especially in the 
conductive regime, we introduce the condensation energy: 
\begin{equation}
E_{\rm c} = E^{\rm F} - E, 
\label{eq:ec}
\end{equation}
where $E^{\rm F}$ denotes the energy per site of the projected 
Fermi sea, $\Psi_Q^{\rm F}={\cal P}\Phi_{\rm F}$, as the reference 
value. 
In Fig.~\ref{fig:cond}, $E^{\rm co}_{\rm c}$, $E^{\rm 120}_{\rm c}$ 
and $E^{d}_{\rm c}$ are shown for three values of $t'/t$. 
Note that $E_{\rm c}$ for every state is almost zero for 
$U<U_{\rm c}^{\rm min}$, where $U_{\rm c}^{\rm min}/t$ is shown 
by an arrow in each panel. 
This means that every state for $U<U_{\rm c}^{\rm min}$ is almost 
reduced to a normal metallic state $\Psi_Q^{\rm F}$. 
Here, it is important to recall that, as discussed in (I),\cite{noteonset}
robust SC occurs only for $U_{\rm onset}^d<U<U_{\rm c}^d$, in which 
$E_{\rm c}/t$ has a small but perceptible finite value. 
Although this tendency can be seen in $E_{\rm c}^d/t$ for $t'/t=0.6$ 
and $6\lsim U/t<7.15$ [Fig.~\ref{fig:cond}(a)], more stable 
$\Psi_Q^{\rm co}$ covers the whole range of SC, namely, 
$U_{\rm c}^{\rm min}=U_{\rm c}^{\rm co}<U_{\rm onset}^d$. 
Consequently, $\Psi_Q^d$ comes to have no chance to arise appreciable 
SC. 
We will return to this subject in \ref{sec:transition}. 
\par

\subsection{\label{sec:transition}Metal-insulator transitions} 
%
In this subsection, we study the critical behavior at $U=U_{\rm c}$ 
found in $E^{\rm co}$ and $E^{\rm 120}$ (cusps) in Fig.~\ref{fig:ene} 
and in $E_{\rm c}^{\rm co}$ and $E_{\rm c}^{\rm 120}$ (sudden increases) 
in Fig.~\ref{fig:cond}. 
Although we have not mentioned, in fact, $E^{\rm co}$ and $E^{\rm 120}$ 
in Fig.~\ref{fig:ene} undergo clear hysteresis (dual-minimum behavior) 
near the cusps at $U_{\rm c}$. 
This indicates a kind of first-order transition takes place at $U_{\rm c}$. 
We will reveal the properties of this transition with various quantities. 
\par

\begin{figure}[!t]
\begin{center}
\includegraphics[width=7.5cm,clip]{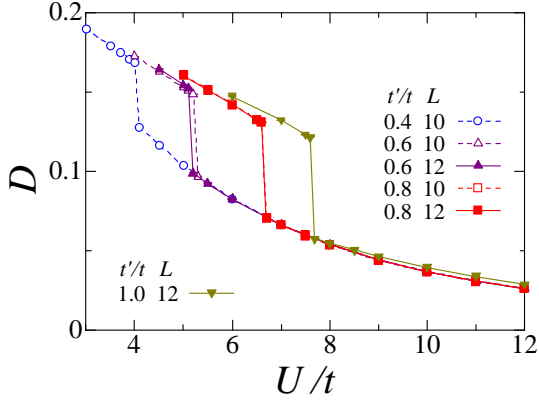}
\end{center}
\vspace{-0.5cm}
\caption{(Color online) 
Density of doubly-occupied site (doublon) as a function of $U/t$ 
for the lowest-energy states: the coexisting state $\Psi_Q^{\rm co}$ 
for $t'/t=0.4$-0.8, and the 120$^\circ$-AF state $\Psi_Q^{\rm 120}$ 
for $t'/t=1.0$. 
}
\label{fig:D} 
\end{figure}
%
First, we take up the doublon density, 
\begin{equation}
D=\frac{1}{N_{\rm s}}\sum_i{\langle n_{i\uparrow}n_{i\downarrow}\rangle}
=\frac{1}{N_{\rm s}}\frac{\langle{\cal H}_{\rm int}\rangle}{U}, 
\end{equation}
where ${\cal H}_{\rm int}$ denotes the second (interaction) term 
of the Hamiltonian eq.~(\ref{eq:model}). 
$D$ is regarded as the order parameter of metal-insulator transitions, 
\cite{Kotliar} by analogy with the particle density in gas-liquid 
transitions. 
As shown in Fig.~\ref{fig:D}, $D$ exhibits a discontinuity 
at $U=U_{\rm c}$ for each $t'/t$, strongly suggesting a first-order 
metal-insulator transition. 
\par

%
\begin{figure}[!t]
\begin{center}
\includegraphics[width=8.0cm,clip]{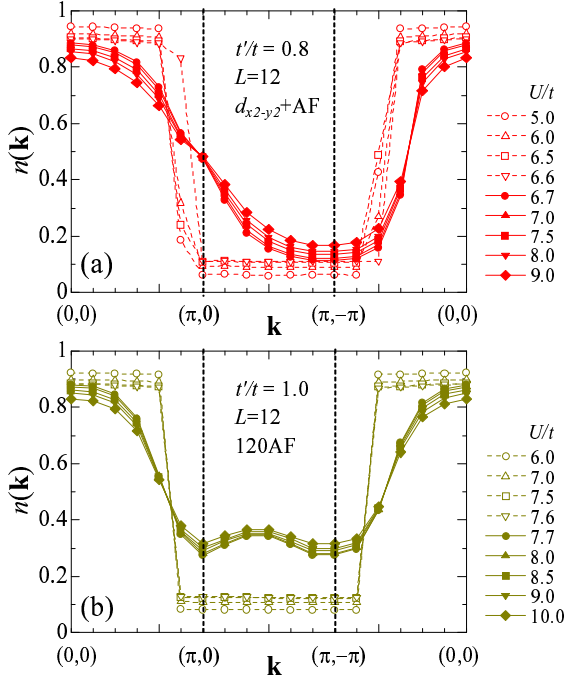}
\end{center}
\vspace{-0.5cm}
\caption{(Color online) 
The momentum distribution function of the lowese-energy 
states is shown for various values of $U/t$ along the path 
$(0,0)$-$(\pi,0)$-$(\pi,-\pi)$-$(0,0)$ in the Brillouin zone,  
(a) for $t'/t=0.8$ (coexisting state $\Psi^{\rm co}_Q$, 
$U_{\rm c}/t \sim 6.65$) and 
(b) for $t'/t = 1.0$ (120$^\circ$-AF state $\Psi^{\rm 120}_Q$, 
$U_{\rm c}/t \sim 7.65$). 
The open (solid) symbols denote the data for $U<U_{\rm c}$ 
($U>U_{\rm c}$). 
}
\label{fig:120nk}
\end{figure}
%
\begin{figure}[!t]
\begin{center}
\includegraphics[width=8.5cm,clip]{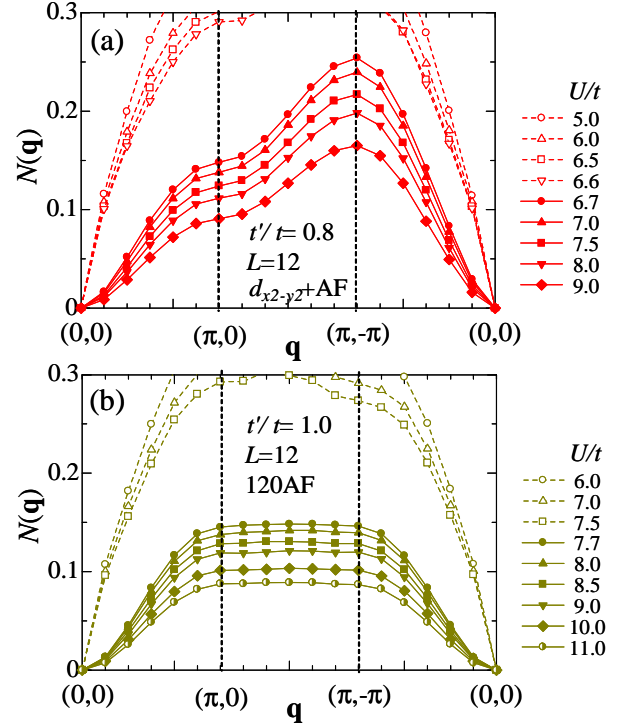}
\end{center}
\vspace{-0.5cm}
\caption{(Color online) 
The charge structure factor $N({\bf q})$ for the same states with 
those in Fig.~\ref{fig:120nk} is plotted along the same path: 
(a) Coexisting state $\Psi_Q^{\rm co}$ for $t'/t=0.8$, and 
(b) 120$^\circ$-AF state $\Psi_Q^{\rm 120}$ for $t'/t=1.0$. 
The open (solid) symbols denote the data for $U<U_{\rm c}$ 
($U>U_{\rm c}$). 
}
\label{fig:nqnq}
\end{figure}
%
\begin{figure}[!t]
\begin{center}
\includegraphics[width=7.5cm,clip]{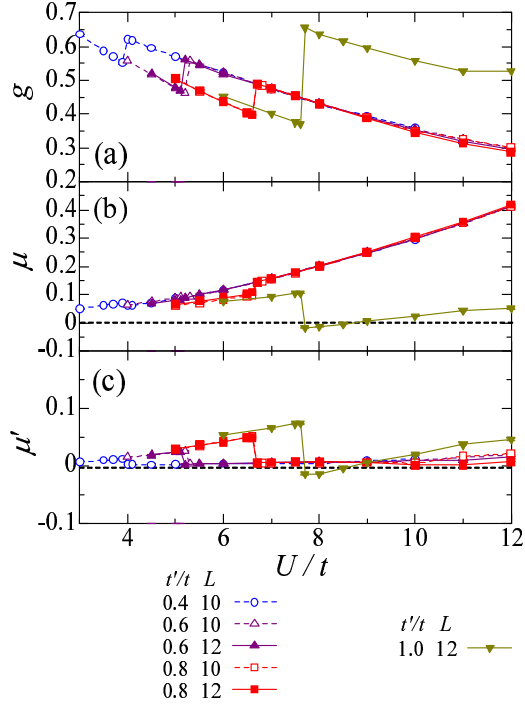}
\end{center} 
\vspace{-0.5cm}
\caption{(Color online) 
Optimized values of variational parameters in correlation factor 
$\cal P$, for several $t'/t$ as function of $U/t$; 
(a) $g$ [onsite (Gutzwiller) correlation parameter], 
(b) $\mu$ [doublon-holon binding parameter in the direction of $t$], 
and (c) $\mu'$ [the same of $t'$]. 
For $t'/t=0.4$-0.8, the parameters are optimized in the coexisting state 
$\Psi_Q^{\rm co}$, and for $t'/t=1.0$, in the 120$^\circ$-AF state 
$\Psi_Q^{\rm 120}$. 
The symbols are common to all panels. 
}
\label{fig:para}
\end{figure}
%
In Fig.~\ref{fig:120nk}, the momentum distribution function, 
\begin{equation}
n({\bf k}) = \frac{1}{2} 
\sum_{\sigma}\langle c_{{\bf k}\sigma}^\dag c_{{\bf k}\sigma}\rangle, 
\end{equation} 
of the lowest-energy states is plotted for $t'/t=0.8$ ($\Psi_Q^{\rm co}$) 
and 1.0 ($\Psi_Q^{\rm 120}$). 
Discontinuities of $n({\bf k})$ at ${\bf k}_{\rm F}$ in both sections, 
$(0,0)$-$(0,\pi)$ and $(0,0)$-$(\pi,\pi)$, are obvious for $U<U_{\rm c}$
for both magnetic states, whereas $n({\bf k})$ becomes smooth in both 
sections for $U>U_{\rm c}$. 
Because the quasi-Fermi surface vanishes for $U>U_{\rm c}$, we may 
consider that the state becomes non-metallic. 
\par

In Fig.~\ref{fig:nqnq}, we depict the charge structure factor, 
\begin{equation}
N({\bf q})=\frac{1}{N_{\rm s}} 
\sum_{i,j}e^{i{\bf q}\cdot({\bf R}_i-{\bf R}_j)} 
\left\langle{N_{i} N_{j}}\right\rangle - n^2, 
\end{equation} 
with $N_{i} = n_{{i}\uparrow} + n_{{i}\downarrow}$, for the same states 
as those in Fig.~\ref{fig:120nk}. 
Similarly to the case of $\Psi_Q^d$ studied in (I), $N({\bf q})$ near 
the $\Gamma$ point $(0,0)$ seems linear in $|{\bf q}|$ for $U<U_{\rm c}$, 
whereas the behaviors of $N({\bf q})$ abruptly change to roughly 
quadratic in $|{\bf q}|$ for $U>U_{\rm c}$, regardless of 
$\Psi_Q^{\rm co}$ or $\Psi_Q^{\rm 120}$. 
It follows that the states are gapless in the charge sector and 
are conductive for $U<U_{\rm c}$, but a charge gap opens for 
$U>U_{\rm c}$ and they become insulating. 
\par

The above results of $D$, $n({\bf k})$ and $N({\bf q})$ indicate 
that in $\Psi_Q^{\rm co}$ and $\Psi_Q^{\rm 120}$, 
a first-order metal-to-insulator transition occurs at $U=U_{\rm c}$, 
as we showed for $\Psi_Q^d$ in (I). 
Nevertheless, the quantities studied below will show that these 
transitions do not belong to pure Mott transitions with no 
relevance to magnetism like in $\Psi_Q^d$, but to metal-to-
magnetic-insulator transitions. 
\par 

Let us consider the optimized variational parameters in the correlation 
factor ${\cal P}$. 
Shown in Figs.~\ref{fig:para}(a)-(c) is the $U/t$ dependence of
the optimized values of $g$, $\mu$ and $\mu'$ for the lowest energy 
states: $\Psi_Q^{\rm co}$ for $t'/t=0.4$-0.8, and $\Psi_Q^{\rm 120}$ 
for $t'/t=1.0$. 
The fact that all the parameters show apparent discontinuities 
at $U=U_{\rm c}$ supports the first-order transition. 
In comparing these values with the corresponding ones for 
$\Psi_Q^d$ shown in Fig.~4 in (I), we notice that the behavior of 
the Gutzwiller parameter $g$ is opposite near the critical point. 
At $U=U_{\rm c}$, $g$ for $\Psi_Q^{\rm co}$ ($t'/t\le 0.8$) becomes 
larger in the insulating side $U>U_{\rm c}$ than in the metallic side 
[Fig.~\ref{fig:para}(a)], 
in contrast to the case for $\Psi_Q^d$ [Fig.~4(a) in (I)].  
This behavior can be understood reasonably, if the $(\pi,\pi)$-AF order 
arises in the insulating regime; it is known \cite{YS2} that $g$ becomes 
larger in a projected ($\pi,\pi$)-AF state than in the corresponding 
paramagnetic state, because the one-body Hartree-Fock state 
$\Phi_{\rm AF}$ already includes an effect to suppress the double 
occupation, in inducing staggered spin structure. 
For $\Psi_Q^{\rm 120}$ ($t'/t = 1.0$), the increase of $g$ at 
$U_{\rm c}$ is still larger than that of $\Psi_Q^{\rm co}$, meaning 
that the triplicate staggered field in $\Phi_{\rm 120}$ forms 
a firmer order for the isotropic case. 
\par

Another noticeable difference is the behavior of the doublon-holon 
binding parameter $\mu$. 
The discontinuity of $\mu$ at $U_{\rm c}$ is an order of magnitude 
smaller in $\Psi_Q^{\rm co}$ than in $\Psi_Q^d$ for $t'/t\le 0.8$. 
This behavior is considered reasonable, again assuming 
the $(\pi,\pi)$-AF order in the insulating regime. 
As we studied before,\cite{YTOT} the doublon-holon binding effect 
is intrinsic in the N\'eel background of $\Phi_{\rm AF}$. 
Accordingly, $\mu$ in the correlation factor ${\cal P}_Q$ plays 
a minor role for the ($\pi,\pi$)-AF state. 
This tendency becomes more thorough for $\Psi_Q^{\rm 120}$; 
$\mu$ in $\Psi_Q^{\rm 120}$, inversely, drops to almost zero 
at $U_{\rm c}$ and remains very small for $U>U_{\rm c}$. 
Similarly, $\mu'$ drops to almost zero at $U_{\rm c}$ for 
$\Psi_Q^{\rm 120}$, and also for $\Psi_Q^{\rm co}$.  
Thus, the doublon-holon binding factor is almost useless for 
$\Psi_Q^{\rm 120}$ in the insulating regime. 
However, in the insulating regime of $\Psi_Q^{\rm 120}$, doublons 
exist as shown in Fig.~\ref{fig:D}, and we have confirmed 
in the records of Monte Carlo sweeps that a doublon almost 
necessarily sits in a nearest-neighbor site of a holon. 
This indicates that the one-body HF state $\Psi_Q^{\rm 120}$ 
already has a sufficient doublon-holon binding effect for finite 
$\Delta_{120}$. 
At any rate, the binding (and unbinding) of a doublon to a holon 
must be the essence of Mott transitions. 
\par

\begin{figure*}[!t]
\begin{center}
\includegraphics[width=14.0cm,clip]{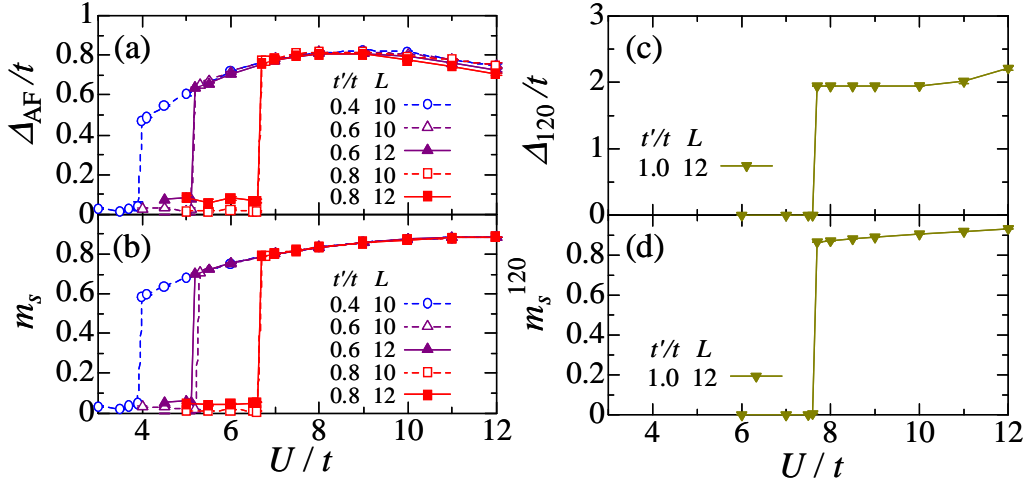}
\end{center} 
\vspace{-0.5cm}
\caption{(Color online) 
(a) Optimized gap parameter $\Delta_{\rm AF}/t$ and 
(b) order parameter $m_{\rm s}$ of a ($\pi,\pi$)-AF order 
for the coexisting state $\Psi_Q^{\rm co}$ ($t'/t=0.4$-0.8). 
(c) Optimized gap parameter $\Delta_{\rm 120}/t$ and 
(d) order parameter $m_{\rm s}^{\rm 120}$ of a 120$^\circ$-AF order 
for the 120$^\circ$-AF state $\Psi_Q^{\rm 120}$ ($t'/t=1.0$). 
For the full polarization, $m_{\rm s}$ and $m_{\rm s}^{\rm 120}$ become 1. 
}
\label{fig:mag}
\end{figure*}

To directly confirm the existence of long-range magnetic orders 
for $U>U_{\rm c}$, we next discuss the behavior of the gap parameters, 
$\Delta_{\rm AF}$ and $\Delta_{\rm 120}$, and the order parameter 
$m_{\rm s}$. 
For $\Psi_Q^{\rm co}$, the sublattice magnetization $m_{\rm s}$ is 
given, as usual, by 
\begin{equation}
m_{\rm s}=\frac{1}{N_{\rm s}}\left|\sum_j e^{i{\bf K}\cdot{\bf R}_j}
          \langle S_j^z\rangle\right|, 
\label{eq:ms}
\end{equation}
with 
$S_j^z = 1/2\left( {c_{j,\uparrow}^\dag c_{j,\uparrow} - 
c_{j,\downarrow}^\dag c_{j,\downarrow}} \right)$. 
Similarly, we define $m_{\rm s}$ for $\Psi_Q^{\rm 120}$ as,  
\begin{equation}
m_{\rm s}^{\rm 120}=\frac{1}{N_{\rm s}}\left|\sum_j e^{i{\bf K}\cdot{\bf R}_j}
          \langle S_j^{Tz}\rangle\right|, 
\label{eq:msT}
\end{equation}
with $S_j^{Tz}=1/2\left( {a_{j,\uparrow}^\dag a_{j,\uparrow} - 
a_{j,\downarrow}^\dag a_{j,\downarrow}} \right)$. 
For $m_{\rm s}^{\rm 120}>0$, $\Psi_Q^{\rm 120}$ has a 120$^\circ$-AF order. 
In Figs.~\ref{fig:mag}(a) and (b), we show $\Delta_{\rm AF}$ and 
$m_{\rm s}$ of $\Psi_Q^{\rm co}$ for three values of $t'/t$ ($\le 0.8$). 
The behavior of these two quantities is similar; 
they are negligibly small for $U<U_{\rm c}$, whereas they abruptly 
increase at $U=U_{\rm c}$ and preserve the large magnitude for $U>U_{\rm c}$. 
They are almost independent of the value of $t'/t$. 
We will turn to this point in \ref{sec:behavior}. 
Shown in Figs.~\ref{fig:mag}(c) and (d) are $\Delta_{\rm 120}$ and 
$m_{\rm s}^{\rm 120}$ of $\Psi_Q^{\rm 120}$ for $t'/t=1.0$. 
Their $U/t$ dependence is basically the same as those of $\Psi_Q^{\rm co}$, 
but the magnitude of $\Delta_{120}/t$ and $m_{\rm s}^{120}$ is 
larger than that of $\Delta_{\rm AF}/t$ and $m_{\rm s}$. 
In this point, the 120-degree AF order is not less steadfast 
than the ($\pi,\pi$)-AF order. 
The spin structure factor $S({\bf q})$ is also checked (not shown), 
which has a sharp peak at ${\bf q}=(2\pi/3,2\pi/3)$ in the insulating regime 
of $\Psi_Q^{\rm 120}$, supporting the realization of the 120$^\circ$ 
spin structure. 
Thus, we have confirmed that a firm magnetic long-range order always 
arises in the insulating regime at least for $t'/t\le 1$. 
\par

%
Finally, we discuss the $d$-wave gap $\Delta_d$ and the $d$-wave 
SC correlation function of the nearest-neighbor-site pairing: 
\begin{eqnarray}
P_d({\bf r})&=&\frac{1}{4N_{\rm s}}
\sum_{i}\sum_{\tau,\tau'=\hat {\bf x},\hat {\bf y}}
(-1)^{1-\delta(\tau,\tau')}\times\qquad \nonumber\\ 
&& \left\langle{\Delta _\tau^\dag({\bf R}_i)\Delta_{\tau'}
({\bf R}_i+{\bf r})}\right\rangle, 
\label{eq:pd}
\end{eqnarray}
where $\hat{\bf x}$ and $\hat{\bf y}$ denote the lattice vectors 
in the $x$ and $y$ directions, and 
$\Delta_\tau^\dag({\bf R}_i)$ is the creation operator of a
nearest-neighbor singlet, 
\begin{equation}
\Delta_\tau^\dag({\bf R}_i)=
(c_{{i}\uparrow}^\dag c_{{i}+\tau\downarrow}^\dag+ 
 c_{{i}+\tau\uparrow}^\dag c_{{i}\downarrow}^\dag)
 /{\sqrt 2}. 
\end{equation}
Unless $\Delta_d$ increases, $P_d({\bf r})$ does not increase, 
but the opposite does not hold, in contrast to the relation between 
$\Delta_{\rm AF}$ and $m_{\rm s}$. 
It is possible that finite $\Delta_d$ indicates a non-SC singlet 
gap.\cite{ZGRS} 
In contrast, $P_d({\bf r})$ is an good indicator of $d_{x^2-y^2}$-wave 
SC, and was studied in detail for $\Psi_Q^d$ in (I), which yielded 
a conclusion that SC arises for $t'/t\lsim 0.7$ within $\Psi_Q^d$. 
Here, we consider the long-distance behavior of $P_d({\bf r})$ by 
$P_d^{\rm ave}$, which is the average of $P_d({\bf r})$ only for 
${\bf r}=(x,L/2)$ and $(L/2,y)$  with $x,y=0$-$L$. 
\par

As shown in Fig.~\ref{fig:pd}(a), $\Delta_d$ for $\Psi_Q^{\rm co}$ is 
always substantially zero for $U<U_{\rm c}$. 
Accordingly, $P_d({\bf r})$ does not develop meaningfully exceeding 
the value of $U=0$, even if $U$ approaches $U_{\rm c}$, as shown in 
Figs.~\ref{fig:pd}(b) and \ref{fig:pd}(c). 
This is in contrast with the case of $\Psi_Q^d$. 
Thus, appreciable SC does not appear in the conducting regime. 
In the insulating regime, the $d$-wave singlet gap $\Delta_d$ is still 
strongly suppressed in $\Psi_Q^{\rm co}$ [Fig.~\ref{fig:pd}(a)], 
compared with in $\Psi_Q^d$ [Fig.~4(c) in (I)], where 
$\Delta_d/t\sim 1.2$-1.3. 
It is found, like the case of $\Psi_Q^d$, $P_d({\bf r})$ is very small 
and vanishes rapidly as $L$ increases (not shown). 
Consequently, for $U>U_{\rm c}$, the ($\pi,\pi$)-AF order is overwhelmingly 
dominant over the $d$-wave SC order; $\Psi_Q^{\rm co}$ in the 
insulating side can be regarded as an almost pure ($\pi,\pi$)-AF 
insulating state. 
It means that $\Psi_Q^{\rm co}$ undergoes a simple first-order 
metal-to-($\pi,\pi$)-AF-insulator transition at $U=U_{\rm c}$ 
\cite{noteorder} for $t'/t\le 0.8$. 
\par

In conclusion, there is no chance that robust $d$-wave SC or 
a nonmagnetic insulator appears within $\Psi_Q^{\rm co}$. 
\par

\subsection{\label{sec:behavior}Antiferromagnetic state and 
phase diagram} 
In this subsection, we consider the properties of the ($\pi,\pi$)-AF 
state realized in the insulating regime of $\Psi_Q^{\rm co}$. 
\par

\begin{figure}[!t]
\begin{center}
\includegraphics[width=7.5cm,clip]{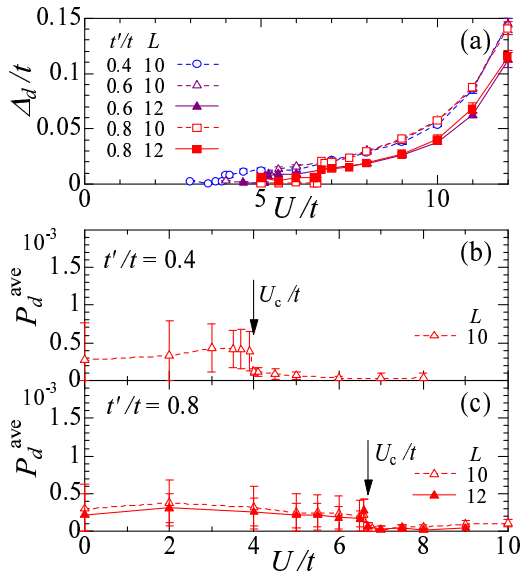}
\end{center} 
\vspace{-0.5cm}
\caption{(Color online) 
(a) Optimized values of $d$-wave gap parameter in $\Psi_Q^{\rm co}$ 
for $t'/t=0.4$-0.8 as a function of $U/t$. 
Averaged nearest-neighbor $d$-wave pairing correlation function in 
$\Psi_Q^{\rm co}$ for (b) $t'/t=0.4$ and (c) $t'/t=0.8$. 
Note that we average $P_d({\bf r})$ only for large values of $|{\bf r}|$ 
(see text). 
For $U/t=0$, we use analytic values. 
The error bars in (b) and (c) include the standard deviations both of 
VMC calculations and by averaging with respect to {\bf r}. 
}
\label{fig:pd}
\end{figure}
%
\begin{figure}[!t]
\begin{center}
\includegraphics[width=7.5cm,clip]{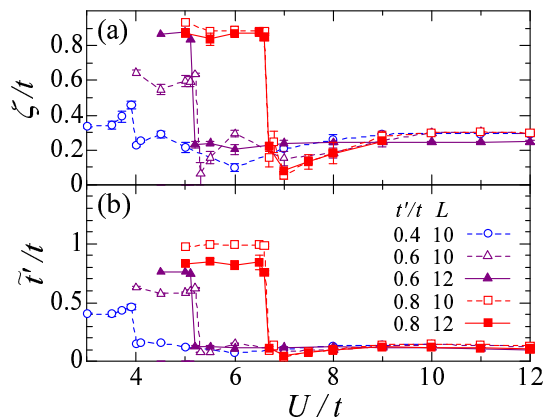}
\end{center} 
\vspace{-0.5cm}
\caption{(Color online) 
Optimized values of the remaining variational parameters in 
$\Psi_Q^{\rm co}$ for $t'/t=0.4$-0.8 as a function of $U/t$; 
(a) $\zeta/t$ [chemical potential], and 
(b) $\tilde{t'}/t$ [band renormalization factor]. 
The symbols are common in both panels. 
}
\label{fig:other}
\end{figure}
%
In (I), we found that the properties of $\Psi_Q^d$ in the (nonmagnetic) 
insulating regime are almost independent of the frustration 
strength $t'/t$ [cf. Fig.~\ref{fig:enetp} for example]. 
This tendency becomes more strong in $\Psi_Q^{\rm co}$. 
As in Fig.~\ref{fig:other}(b), the renormalized frustration $\tilde t'/t$ 
becomes nearly zero for $U>U_{\rm c}$, regardless of the model parameter 
$t'/t$, namely, in the strong coupling regime, the effective band almost 
retrieves the nesting condition for the simple square lattice ($t'=0$), even 
for highly frastrated cases.\cite{reno} 
The other variational parameters in $\Psi_Q^{\rm co}$ are also almost 
independent of $t'/t$, as seen in each panel of Figs.~\ref{fig:para}, 
\ref{fig:mag}(a), \ref{fig:pd}(a) and \ref{fig:other}(a), where all 
the data points for $U>U_{\rm c}$ are represented very well by a unique 
curve, regardless of $t'/t$. 
Thus, the optimized $\Psi_Q^{\rm co}$ is not changed with the frustration 
strength, as long as $U>U_{\rm c}$. 
\par

\begin{table}
\caption{
Energy components and total energy of $\Psi_Q^{\rm co}$ 
for three values of $U/t$ in the regime of the ($\pi,\pi$)-AF insulator 
($U>U_{\rm c}$). 
Here, $t'/t=0.8$ ($U_{\rm c}/t\sim 6.65$). 
The small system-size dependence is a characteristic of an ($\pi,\pi$)-AF 
state. \cite{YS2}
The digits in brackets indicate the errors in the last digits. 
}
\vspace{1mm}
\label{table:1}
\begin{tabular}{c|c|c|c|c|c} \hline
 $U/t$ & $L$ & $E_t/t$    & $E_{t'}/t$   & $E_U/t$   & $E/t$      \\
\hline
  10   &  10 & -0.7761(6) & -0.0001(0) & 0.3659(6) & -0.4103(1) \\
       &  12 & -0.7759(9) & -0.0001(0) & 0.3657(9) & -0.4103(1) \\
\hline
  12   &  10 & -0.6618(7) & -0.0002(0) & 0.3134(7) & -0.3485(1) \\
       &  12 & -0.6601(6) & -0.0001(0) & 0.3119(6) & -0.3483(1) \\
\hline
  14   &  10 & -0.5749(5) & -0.0002(0) & 0.2713(6) & -0.3038(1) \\
       &  12 & -0.5738(5) & -0.0001(0) & 0.2703(7) & -0.3035(1) \\
\hline
\end{tabular}
\end{table}
%
In Fig.~\ref{fig:enetp}, the total energy for $\Psi_Q^{\rm co}$ 
in the insulating regime ($U/t=12$) is plotted as a function of $t'/t$. 
Here, $E^{\rm co}$ is almost constant, and the difference of $E/t$ 
between $t'/t=0$ and 1.2 is as small as 0.1\%. 
This behavior is not trivial even if the wave function is not changed 
with $t'/t$, because the $t'$-term in the Hamiltonian changes. 
To understand this result, we check the behavior of energy components; 
let $E_t$, $E_{t'}$ and $E_U$ be the contributions from the hopping in 
the $t$-bond and $t'$-bond directions, and from the onsite interaction 
$U$, respectively. 
We list the numerical data for $t'/t=0.8$ in Table \ref{table:1} 
as a typical example, because each contribution is again almost 
constant as a function of $t'/t$. 
As expected, $E_{t'}$ is substantially zero, indicating if we allow 
the band renormalization, the wave function is by far stabilized by 
retrieving the nesting condition for the simple square lattice 
at the cost of the energy reduction due to the diagonal hopping or 
frustration, even if $t'/t$ is considerably large. 
\par

It is natural to guess that this renormalization readily occurs in 
$\Psi_Q^{\rm co}$, because the nesting condition is 
advantageous not only to the ($\pi,\pi$)-AF state but to the $d$-wave state, 
as discussed in (I). 
Anyway, in recalling the point (iv) itemized in \ref{sec:method}, 
we notice that the band renormalization effect, namely the recovery 
of nesting, is essential to stabilize the ($\pi,\pi$)-AF state, 
as well as the $d$-wave singlet state. \cite{YOT}
\par

\begin{figure}
\begin{center}
\includegraphics[width=8.0cm,clip]{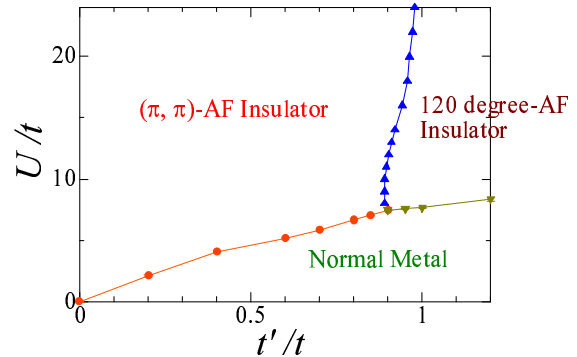}
\end{center}
\vspace{-0.5cm}
\caption{(Color online) 
Ground-state phase diagram in the $t'$-$U$ plane constructed from the 
present VMC results of the coexisting wave function $\Psi_Q^{\rm co}$ 
and the 120$^\circ$-AF state $\Psi_Q^{\rm 120}$. 
At the boundaries of the metallic and insulating phases, 
first-order magnetic transitions take place. 
}
\label{fig:phased}
\end{figure}
%
Finally, we discuss the ground-state phase diagram, which is 
reconstructed within $\Psi_Q^{\rm co}$ and $\Psi_Q^{\rm 120}$ and 
depicted in Fig.~\ref{fig:phased}. 
As compared with the diagram by $\Psi_Q^d$ and $\Psi_Q^{\rm AF}$ 
shown in Fig.~14 in (I), the area of the ($\pi,\pi$)-AF insulator 
extends to extremely large $t'/t$ ($>0.9$) and to somewhat small $U/t$. 
In addition, the area of the 120$^\circ$-AF insulator appears 
near the isotropic point $t'/t=1$. 
We consider these tendencies are broadly consistent with the results 
for the $J$-$J'$ model ($U/t=\infty$), \cite{J-J'} in which 
the domain of ($\pi,\pi$)-AF continues to $t'/t>0.8$. 
In Fig.~\ref{fig:phased}, as $U/t$ increases, the boundary value in 
$t'/t$ between the ($\pi,\pi$)-AF and 120$^\circ$-AF insulators 
tends to increase. 
This is probably because $\Psi_Q^{\rm co}$ is stabilized by the 
$d$-wave gap $\Delta_d$, which rapidly increases for large $U/t$,
as seen in Fig.~\ref{fig:pd}(a). 
We consider that the above tendency of the boundary will be corrected 
by introducing an appropriate singlet gap also into $\Psi_Q^{\rm 120}$. 
As a result of the stabilization of magnetic phases, the domains of 
nonmagnetic insulating and of robust $d$-wave SC phases disappear, 
which occupy certain parts of the phase diagram made in (I) and 
also in recent studies of a variational cluster perturbation 
theory \cite{V-CPT} and a cellular dynamical mean field theory. \cite{CDMFT}
\par

\subsection{\label{sec:extd}Extention of pairing-gap form} 

\begin{figure}[!t] 
\begin{center}
\includegraphics[width=8.0cm,clip]{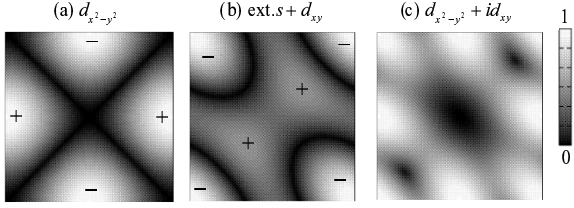}
\end{center}
\caption{
Magnitude of pairing potentials $|\Delta_{\bf k}/\Delta_{\rm max}|$ 
considered in BCS state: 
(a) $d_{x^2-y^2}$, 
(b) ${\rm ext}.s+d_{xy}$, and 
(c) $d_{x^2-y^2}+id_{xy}$. 
$|\Delta_{\rm max}|$ denotes the maximum of $|\Delta_{\bf k}|$ for 
each pairing gap. 
}
\label{fig:gapfun}
\end{figure}
From the argument in \ref{sec:behavior}, we expect a state yielding 
a gain in $E_{t'}$ overcomes $\Psi_Q^{\rm co}$ and $\Psi_Q^{\rm 120}$ 
for large $t'/t$. 
In this subsection, we consider a couple of different pairing gaps, 
which seem suitable for $t'\gsim t$, in the projected BCS function. 
\par

One has a specific gap parameter to the $t'$ direction ($\Delta_{d'}$), 
independent of $\Delta_s$ for the $t$ direction, \cite{Tanuma,Liu} 
\begin{equation}
\Delta_{\bf k} = \Delta_{s}(\cos k_x + \cos k_y)-
\Delta_{d'} \cos (k_x+k_y), 
\label{eq:ext-d}
\end{equation}
which we call ``${\rm ext}.s$+$d_{xy}$ wave" 
($\Psi_Q^{s+d'}={\cal P}\Phi_{s+d'}$). 
This form of $\Delta_{\bf k}$ has nodes near the $k_x$ and $k_y$ axes 
for $\Delta_{s} \sim \Delta_{d'}$ [see Fig.~\ref{fig:gapfun}(b)], 
which resembles the nodes proposed by some experiments. \cite{Izawa,Arai} 
$\Delta_{\bf k}$ approaches the $d_{xy}$ wave of a one-dimensional 
character for $|\Delta_{d'}|\gg |\Delta_s|$. 
The other is a $d_{x^2-y^2}$+$id_{xy}$ wave 
($\Psi_Q^{d+id}={\cal P}\Phi_{d+id}$), 
\begin{equation}
\Delta_{\bf k} \!=\! \Delta_{d+id} \!
\left[{\cos k_x \!+\! e^{i\frac{{2\pi}}{3}}\! \cos(k_x \!+\! k_y) 
\!+\! e^{i\frac{{4\pi}}{3}}\! \cos k_y}\right], 
\label{eq:d+id}
\end{equation}
as shown in Fig.~\ref{fig:gapfun}(c). 
This form was often used to study favorable gap symmetries for 
cobaltate SC; \cite{sbmf,OgataGA,WataCo} using a VMC method \cite{WataCo} 
for the $t$-$J$ model on an isotropic triangular lattice, it was shown 
that $\Psi_Q^{d+id}$ is degenerate with $\Psi_Q^d$ at half filling, 
and has lower energy for doped cases. 
This gap form breaks a time reversal symmetry. 
\par

In Fig.~\ref{fig:enetp}, the total energies of $\Psi_Q^{s+d'}$ 
($E^{s+d'}$) and $\Psi_Q^{d+id}$ ($E^{d+id}$) are plotted in addition 
to those mentioned earlier. 
For $t'/t\lsim 1.1$, $E^{d+id}$ is almost constant in the same reason 
as $E^{\rm co}$ and $E^d$, whereas $E^{d+id}$ starts to decreases 
at $t'/t\sim 1.1$ abruptly, because, there, the direction of band 
renormalization is reversed from $\tilde t'/t\rightarrow 0$ to 
$\tilde t'/t\rightarrow\infty$. 
Thus, the effective Fermi surface of $\Psi_Q^{d+id}$ becomes quasi 
one dimensional for $t'/t\gsim 1.1$. 
Similarly to $E^{d+id}$, $E^{s+d'}$ considerably decreases as $t'/t$ 
increases. 
In the range of decreasing $E$, the energy reduction in both 
$\Psi_Q^{d+id}$ and $\Psi_Q^{s+d'}$ is largely attributed to $E_{t'}$. 
Especially in $\Psi_Q^{s+d'}$, the energy reduction is entirely 
owing to $E_{t'}$, and the direction of band renormalization is 
$\tilde t'/t\rightarrow\infty$; the optimized $\Delta_{s}$ is 
negligible ($\sim 0.54$) compared to the optimized $\Delta_d'$ ($\sim 7.05$), 
for $U/t=12$, $t'/t=1.2$, and $L=12$. 
Thus, the singlet gap has an almost pure $d_{xy}$-wave of one-dimensional 
character. 
As shown in Fig.~\ref{fig:enetp}, 
$E^{s+d'}$ overcomes $E^{120}$ for $t'\gsim t'_{\rm c2}\sim 1.65t$
for $U/t=12$, 
meaning that $\Psi_Q^{\rm 120}$ is predominant for an unexpectedly 
large range of $t'/t$ $(>1)$ within the states we have studied 
($L=10$ and $12$). 
We expect a more favorable pairing gap will be found for 
$t<t'<t'_{\rm c2}$, but we leave a search for it for the future. 
\par

Detailed results for $\Psi_Q^{s+d'}$ was reported in another 
publication. \cite{WataOD} 
\par

\section{Conclusion\label{sec:conclusion}}
\subsection{\label{sec:summary}Summary}
As a continuation of the preceding study (I), \cite{Wata} we have studied 
the Hubbard model on anisotropic triangular lattices, eq.~(\ref{eq:model}), 
at half filling, using an optimization variational Monte Carlo method. 
We introduce two new trial wave functions: 
(i) A coexisting state of ($\pi,\pi$)-AF and $d$-wave gaps, which 
allows for a band renormalization effect, $\Psi_Q^{\rm co}$, and 
(ii) a state with an AF order of 120$^\circ$ spin structure, 
$\Psi_Q^{\rm 120}$. 
Main results are summarized as follows: 
\par

[1] First-order metal-to-insulator transitions occur in both 
$\Psi_Q^{\rm co}$ and $\Psi_Q^{\rm 120}$ at smaller values of $U/t$ than 
those of the $d$-wave state $\Psi_Q^d$ studied in the preceding paper (I). 
As a result, the regime of robust $d$-wave SC found in (I) is covered with 
the domain of these states. 
The modified phase diagram within $\Psi_Q^{\rm co}$ and $\Psi_Q^{\rm 120}$ 
is shown in Fig.~\ref{fig:phased}. 

[2] In the insulating regimes, $\Psi_Q^{\rm co}$ and $\Psi_Q^{\rm 120}$ 
are considerably stable, compared with $\Psi_Q^d$, and magnetic long-range 
orders always exist for $t'/t\lsim 1.65$. 
Thus, a domain of a nonmagnetic insulator is not found for $t'\sim t$ 
within the wave functions used this time. 
\par

[3] In the insulating regime of $\Psi_Q^{\rm co}$, the realized state 
can be regarded as a pure ($\pi,\pi$)-AF insulator, because the sublattice 
magnetization as well as the ($\pi,\pi$)-AF gap ($\Delta_{\rm AF}$) is robust, 
and the $d$-wave pairing correlation almost vanishes. 
In the optimized $\Psi_Q^{\rm co}$, the effective band is renormalized 
so greatly ($\tilde t'\rightarrow 0$), irrespective of $t'/t$, that 
the nesting condition for $t'=0$ is retrieved almost completely. 
Accordingly, the contribution of diagonal hopping energy vanishes 
even for large $t'/t$. 
\par

[4] For $t'\sim t$, $\Psi_Q^{\rm 120}$ becomes predominant ($U>U_{\rm c}$), 
even though the effects of band renormalization and of coexisting singlet 
gaps are not considered. 
If these effects are introduced, the area of the 120$^\circ$-AF order 
will somewhat expands, although, at present, the area of 
the ($\pi,\pi$)-AF order extends to as large as $t'/t\gsim 0.9$. 
\par

[5] For large values of $t'$ ($>t_{\rm c2}\sim 1.65$), the singlet pairing 
states with gaps oriented to the diagonal-bond direction overcome 
$\Psi_Q^{\rm 120}$. 
We speculate that another predominant singlet (and SC) state will be 
discovered for $t<t'<t_{\rm c2}$.
\par

We believe that the mechanisms of a Mott (conductive-to-nonmagnetic 
insulator) transition and of the $d_{x^2-y^2}$-wave SC pursued in
(I) fundamentally remain valid, if the magnetic orders are removed 
for some reasons. 
However, the ground-state phase diagram for the model eq.~(\ref{eq:model}) 
is substantially modified by $\Psi_Q^{\rm co}$ and $\Psi_Q^{\rm 120}$. 
\par

\subsection{Discussions\label{sec:discussions}}
In comparing the present results with experimental ones of $\kappa$-ET 
salts, a favorable point is that a ($\pi,\pi$)-AF insulator is realized 
for realistic values of $t'/t$, namely e.g. $0.74$ 
in $\kappa$-(ET)$_2{\rm Cu{N(CN)_2}Cl}$. 
An unfavorable point is that robust SC and a nonmagnetic insulator 
do not appear; the latter state is believed to be realized in 
$\kappa$-(ET)$_2$Cu$_2$(CN)$_3$.\cite{Shimizu} 
One conceivable cause of this discrepancy is the insufficiency 
of trial wave functions; it is possible that quantum fluctuation 
is not sufficient for $U\sim U_{\rm c}$ and large $t'/t$, and that 
we have not exhausted crucial orders. 
Another possible cause is that the present model eq.~(\ref{eq:model}) 
is not sufficient to describe $\kappa$-ET salts. 
For instance, the dimerization of ET molecules is not strong 
enough to justify the use of a single-band model. \cite{Kuroki}
\par

In the theoretical point of view, the present result is comparable 
to that for $U/t\rightarrow\infty$, namely the $J$-$J'$ Heisenberg 
model. 
According to it, the ($\pi,\pi$)-AF long-range order vanishes 
at $t'/t\sim 0.8$, \cite{J-J'} and an AF order with 120$^\circ$ 
spin structure prevails at $t'/t=1$,\cite{tri} although a disordered 
phase may intervene between the two magnetic phases. 
Some other theoretical studies \cite{PIRG,V-CPT,CDMFT,Koretsune} 
for the equivalent Hubbard model have yielded results of nonmagnetic 
insulating states at $t'/t\sim 1$. 
However, these studies have not explicitly treated the 120$^\circ$-AF 
order, which is shown very stable for $t'/t = 1$ in this study. 
\par

Although robust SC does not appear within the present study, we 
found that the symmetry of a singlet gap changes at large $t'/t$ ($\sim 1.2$) 
from the simple $d_{x^2-y^2}$ wave to, for instance, the $d_{xy}$ wave 
as mentioned in \ref{sec:extd} (see Fig.~\ref{fig:enetp}). 
This aspect is in accordance with that of FLEX, \cite{Moriya} 
in which a predominant SC symmetry switches from a $d_{x^2-y^2}$-wave 
to a $d_{xy}$-wave state at $t'/t\sim 1$. 
Owing to this competition between $d_{x^2-y^2}$ and $d_{xy}$ waves 
near the isotropic point ($t'/t=1$), the SC gap symmetry realized 
in $\kappa$-ET salts, especially in $\kappa$-(ET)$_2$Cu$_2$(CN)$_3$, 
may not be definitive but sensitive to physical parameters such 
as pressure. 
In contrast, a recent study of the Hubbard model with an extra exchange 
term using a Gutzwiller approximation \cite{Gan} concluded 
that a $d$+$id$-wave SC is stable for $U\sim W$ and $t'\gsim t$. 
Thus, it is urgent to carry out VMC calculations, in which the form 
of the pairing gap can be optimized without biased assumptions. 
\par

\begin{acknowledgments}
The authors appreciate the useful communication with Yung-Chung Chen, 
who has independently pointed out the importance of the renormalization 
of $\varepsilon_{\bf k}$ for the AF phase. \cite{Chen} 
The authors thank Masao Ogata and Kenji Kobayashi for useful discussions. 
This work is partly supported by Grant-in-Aids from the Ministry of 
Education, etc. Japan, from the Supercomputer Center, ISSP, Univ. of Tokyo, 
from NAREGI Nanoscience Project and for the 21st Century COE "Frontiers 
of Computational Science". 
\end{acknowledgments} 
\par




\begin{thebibliography}{99}
\bibitem{noteonset} 
As studied in (I), $E_{\rm c}^d$ for small $t'/t$ ($\lsim 0.7$) starts 
to increase gradually at $U\sim U_{\rm onset}^d$ ($<U_{\rm c}^d$), 
as seen also in Fig.~\ref{fig:cond}(a) (in this case 
$U_{\rm onset}^d/t\sim 6$). 
This increase stems from the SC gap opening. 
On the other hand, such increase cannot be seen in $E_{\rm c}^{\rm co}$, 
for arbitrary $t'/t$. 
This strongly suggests that robust SC is unlikely to arise in 
$\Psi_Q^{\rm co}$.
\bibitem{noteorder} For $t'=0$, a continuous metal-to-AF-insulator 
transition takes place at $U=0$. 
It follows that the character of the transition changes from continuous 
to first-order at $t'\sim 0$.
\bibitem{reno} In some weak-correlation approaches \cite{Ogawa}, 
band renormalization toward the direction of retrieving the nesting condition 
has been found $in$ $the$ $conducting$ $regimes$. 
However, these are phenomena essentially different from the very large 
band renormalization $in$ $the$ $insulating$ $regime$ discussed in this paper. 
As seen in Fig.~\ref{fig:other}(b), the renormalization for $U<U_{\rm c}$ 
is too small to detect in the present study. 
Incidentally, similar large band renormalization in strongly-correlated 
regimes ($U>U_{\rm c}$) has been found for $d$-wave states \cite{Liu,Wata}.
\end{thebibliography}
\end{document}